\documentclass[]{emulateapj}

\begin{document}

\title{Asteroseismology of the pre-main-sequence $\delta$ Scuti pulsator IP Persei}
\author{Xinghao Chen\altaffilmark{1,2} and Yan Li \altaffilmark{1,2,3,4}}

\altaffiltext{1}{Yunnan Observatories, Chinese Academy of Sciences, P.O. Box 110, Kunming 650216, China; chenxinghao@ynao.ac.cn; ly@ynao.ac.cn}
\altaffiltext{2}{Key Laboratory for Structure and Evolution of Celestial Objects, Chinese Academy of Sciences, P.O. Box 110, Kunming 650216, China}
\altaffiltext{3}{University of Chinese Academy of Sciences, Beijing 100049, China}
\altaffiltext{4}{Center for Astronomical Mega-Science, Chinese Academy of Sciences, 20A Datun Road, Chaoyang District, Beijing, 100012, China}
\begin{abstract}
A grid of theoretical models are computed to fit the 9 oscillation modes of IP Per detected earlier from a multi-site ground-based campaign. Fitting results show that there are two sets of theoretical models that could reproduce the observed frequencies more or less euqally well. In view of other available spectroscopic and photometric measurements, our best fitting stellar parameters for IP Per are $\upsilon_{\rm e} = 91^{+5}_{-3}$ km s$^{-1}$, $Z=0.009^{+0.004}_{-0.001}$, $M = 1.64^{+0.10}_{-0.04}$ $M_{\odot}$, $T_{\rm eff} = 7766^{+348}_{-173}$ K, $\log L/L_{\odot} = 1.125^{+0.094}_{-0.046}$, $\log g = 4.041^{+0.008}_{-0.003}$ dex, $R = 2.022^{+0.042}_{-0.018}$ $R_{\odot}$,  $\tau_{0} = 8711^{+68}_{-35}$ s, age = $7.39^{+0.76}_{-0.46}$ Myr. Meanwhile, IP Per is found to be a pre-main sequence star where CN cycle has not yet reached the equilibrium state. At present fourteen percent of C12 have been turned into N14. Based on the best-fitting model, $f_{6}$ is identified as a radial mode, $f_{1}$ and $f_{2}$ as two dipole modes, and $f_{3}$, $f_{4}$, $f_{5}$, $f_{7}$, $f_{8}$, and $f_{9}$ as six quadrupole modes.
\end{abstract}


\section{Introduction}
The $\delta$ Scuti pulsating stars are a class of A- and F-type stars situated at the position where the main sequence overlaps with the classical instability strip. Their pulsation periods range from 0.5 to 6 hours (Breger 2000). Thanks to the space missions MOST (Walker et al. 2003), CoRoT (Baglin et al. 2006), and Kepler (Borucki et al. 2010), a lot of $\delta$ Scuti stars have been observed precisely and a large amount of oscillation modes are obtained (e.g., HD 144277 (Zwintz et al. 2011), CoRoT 102749568 (Papar$\acute{\rm o}$ et al. 2013), and HIP 80059 (Ripepi et al. 2015)). Both of radial and nonradial pulsations are detected, thus $\delta$ Scuti pulsating stars are very important and promising objects for asteroseismology.

However, mode identifications are very difficult for the $\delta$ Scuti stars, because the asymptotic theory of nonradial oscillations (Tassoul et al. 1980; Unno et al. 1989) is invalid for low-order modes. Two common methods of mode identifications for $\delta$ Scuti stars are the spectroscopic method (Mantegazza 2000) and the photometric method (Watson 1988), respectively. Due to the complexity of the frequency spectra of $\delta$ Scuti stars, only a few modes can be identified by using the two methods. Besides, Papar$\acute{\rm o}$ et al. (2016a, 2016b) developed a method of sequence search, and found a large amount of independent series of regular frequency spacing in 77 $\delta$ Scuti stars. Moreover, Chen et al. found that frequency spectra of $\delta$ Scuti stars can be disentangled using the method of rotational splitting, such as HD 50844 (Chen et al. 2016), CoRoT 102749568 (Chen et al. 2017), and EE Cam (Chen $\&$ Li 2017).

The $\delta$ Scuti stars are mainly in the evolutionary stage of the main sequence and post-main sequence (Breger et al. 2000), but also in the stage of pre-main sequence (Breger 1972; Zwintz 2008; Zwintz et al. 2011; Zwintz et al. 2014). Because evolutionary tracks of pre-main sequence stage usually intersect with those of main sequence stage and post-main sequence stage on the Hertzsprung-Russel diagram, it is hard to distinguish them from their more evolved counterparts. Besides, obtaining reliable masses and ages of the pre-main sequence stars is also very hard. Reliable masses are essential to investigate the initial mass function and its possible dependence on the features of the parent cloud. We have performed a detailed asteroseismic study for the $\delta$ Scuti star HIP 80088 (Chen $\&$ Li 2018). We precisely determined the evolutionary stage of HIP 80088, i.e., the CN cycle has not reached the equilibrium state (twenty-eight percent of C12 have been turned into N14). Meanwhile, the mass of HIP 80088 is also determined precisely. That motivates us to analyse another $\delta$ Scuti pulsating star with asteroseismology. The pulsation frequencies of IP Per range from 107.64 to 601.85 $\mu$Hz (referring to Table 1), covering a wide frequency range, thus we choose the $\delta$ Scuti star IP Per as our research object in this work.

IP Per is classified as an A-type star (Wenzel 1978; Gray $\&$ Corbally 1998), with apparent magnitude 10.374 in Johnson V (Ripepi et al. 2006). Its variability was discovered by Hoffmeister (1949), and found to be similar to those of Herbig Ae stars (Grinin er al 1991). Herbig Ae stars are a class of pre-main sequence intermediate mass objects discovered about 60 years ago (Herbig 1960). They are well known for irregular short-term brightness minima accompanied by an increase of the optical reddening. The catalogue of Th$\acute{\rm e}$ et al. (1994) classified IP Per as a pre-main sequence intermediate mass star. Miroshnichenko et al. (2001) found significant variations of the Balmer line profiles and a large IR excess for IP Per, which are in agreement with the stage of pre-main sequence.
Besides, de Zeeuw et al. (1999) identified IP Per as a member star of Perseus OB2 association based on the radial velocity and the proper motion. Perseus OB2 association is the second closest OB association to the Sun,  with an age less than 15 Myr (de Zeeuw et al. 1999; Bally et al. 2008; Azimlu et al. 2015).

Kovalchuk $\&$ Pugach (1997) investigated the surface gravitational accelerations of 19 variable Herbig Ae/Be stars, and found that their values of $\log g$ are significantly lower than those expected for main sequence stars with similar spectral types. They obtained that $T_{\rm eff}$ and $\log g$ of IP Per are 8800 K and 2.0 respectively. Soon after, Miroshnichenko et al. (2001) carried out a detailed comparison between the spectra of IP Per and those of other stars, as well as theoretical models. They derived IP Per as an A7V star with $T_{\rm eff} = 7960 \pm 50$ K, $\log g = 4.38 \pm 0.04$, [M/H] = $-0.41\pm 0.04$, $\upsilon$sin$i = 70 \pm 20$ km s$^{-1}$. Besides, Zwintz et al. (2014) determined stellar parameters of IP Per to be $T_{\rm eff} = 8000 \pm 100$ K, $\log g= 4.40 \pm 0.10$, [Fe/H] = $-0.24$, and $\upsilon$sin$i = 70 \pm 4$ km s$^{-1}$ using the high-resolution spectroscopy from McDonald Observatory Tull spectrograph.

In addition, Ripepi et al. (2006) obtained a total of about 190 hours photometric data over 38 nights on the basis of a photometric multisite campaign on IP Per. They analyzed the three datasets 2002 B, 2003 B, and 2003 V respectively, and obtained 9 independent frequencies ranging from 107 to 602 $\mu$Hz. The dataset 2003 V is the longest, and yields the most number of oscillation frequencies. Table 1 lists the nine frequencies of the dataset 2003 V. The uncertainties of these frequencies are $\pm$0.11 c/d. Moreover, Ripepi et al. (2006) compared the observed frequencies with theoretical models, and found that they could not reproduce all observed modes for their selected stellar parameters. We extend the work of Ripepi et al. (2006) and perform a more comprehensive asteroseismic analysis for IP Per. In Section (2), we introduce our input physics and model grids. In Section (3), we present our asteroseismic analysis on the $\delta$ Scuti pulsator IP Per. We describe details of model fittings in Section 3.1, elaborate the fitting results in Section 3.2, and discuss them in Section (3.3). Finally, we summarize the main results in Section 4.

\section{Input physics and model grids}
In our work, theoretical models are computed with the Modules for Experiments in Stellar Astrophysics (MESA), which is one-dimensional open source stellar evolution code developed by Paxton et al. (2011, 2013, 2015, 2018). We use the submodule "pulse" of version 6596 to construct stellar evolutionary models (Paxton et al. 2011, 2013) and to calculate frequencies of their corresponding oscillation modes (Christensen-Dalsgaard 2008).

In our calculations, we choose the 2005 update of the OPAL equation of state  (Rogers $\&$ Nayfonov 2002). In the high temperature region, we use the OPAL opacity tables from Iglesias $\&$ Rogers (1996). In the lower temperature region, we use opacity tables from Ferguson et al. (2005) instead. We use the solar metal composition GS98 (Grevesse $\&$ Sauval 1998) as the initial composition in metal. For the atmosphere boundary conditions, we adopt the default setting of MESA. In addition, we use the mixing-length theory (MLT) of B$\ddot{\rm o}$hm-Vitense (1958) to treat convection. Finally, we do not consider effects of rotation, element diffusion, and the convective overshooting on the structure and evolution of a star in this work.

A grid of stellar models are constructed with MESA. Each star evolves from the pre-main sequence to the age of the star arriving at 200 Myr. When the age of the star is less than 20 Myr, the  maximum time step is set to be  500 years. When the age of the star is larger than 20 Myr, the maximum time step is set to be 0.1 Myr. Thus, the step size in $T_{\rm eff}$ is always less than 1 K, and the step size in $\log g$ is always less than 0.0002 dex. The calibrated value of the mixing- length parameter $\alpha = 1.8$ for the sun is adopted. We follow the works of Dotter et al. (2008) and Thompson et al. (2014), and adopt the initial helium fraction $Y=0.245+1.54Z$ as a function of the mass fraction of heavy-elements $Z$. Then the structure and evolutionary track of a star are determined by the stellar mass $M$ and the mass fraction of heavy-elements $Z$. The mass $M$ ranges from 1.5 $M_{\odot}$ to 2.5 $M_{\odot}$ with a step of 0.02 $M_{\odot}$, and $Z$ ranges from 0.005 to 0.030 with a step of 0.001.

 \section{Asteroseismic anaylsis}
\subsection{Scheme of Model fittings}
The theory of stellar oscillation shows that an oscillation mode can be characterized by three indices ($n$, $\ell$, $m$). Therein, $n$ is the radial order, $\ell$ is the spherical harmonic degree, and $m$ is the azimuthal number. The azimuthal number $m$ is degenerate for a spherically symmetric star, i.e., frequencies of oscillation modes with the same $n$ and $\ell$ but different $m$ are the same. If a star is rotating, effects of rotation will break the spherical symmetry of the star, and will split one nonradial oscillation mode into $2\ell+1$ different ones. The general formula of the first-order effect of rotation is deduced to be 
\begin{equation}
\nu_{\ell,n,m}= \nu_{\ell,n} + m\delta\nu_{\ell ,n} =  \nu_{\ell,n} + \beta_{\ell, n}\frac{m \upsilon_{\rm e}}{2\pi R} ,
\end{equation}
(Aerts et al. 2010). In equation (1), $\delta\nu_{\ell ,n} $ is the splitting frequency, $R$ is the stellar radius, $\upsilon_{\rm e}$ is the equatorial rotational velocity, and $\beta_{\ell,n}$ is the rotational parameter. The effect of rotation can be completely given by the constant $\beta_{\ell,n}$. Its general expression for a uniformly rotating star can be described as 
\begin{equation}
\beta_{\ell, n}=\frac{\int_{0}^{R}(\xi_{r}^{2}+L^{2}\xi_{h}^{2}-2\xi_{r}\xi_{h}-\xi_{h}^{2})r^{2}\rho dr}
{\int_{0}^{R}(\xi_{r}^{2}+L^{2}\xi_{h}^{2})r^{2}\rho dr}
\end{equation}
(Aerts et al. 2010), where $\xi_{r}$ is the radial displacement, $\xi_{h}$ is the horizontal displacement, $\rho$ is the local density, and $L^{2}= \ell(\ell+1)$.

Figure 1 depicts evolutionary tracks of stars with $M$ ranging from 1.50 to 2.50 $M_{\odot}$ in a step of 0.10 $M_{\odot}$, and $Z$ ranging from 0.005 to 0.030 in a step of 0.005. According to our calculations, we find that theoretical models with $\log g = 4.4 \pm 0.1$ can not reproduce the observed frequency spectra of IP Per. We then adopt a wide observational constraints, i.e.,  7500 K $< T_{\rm eff} < $ 8500 K and $4.0 < \log g < 5.0$. We mark the constraints in Figure 1 with one rectangle. When a star evolves along its evolutionary track, we calculate frequencies of oscillation modes with $\ell$ = 0, 1, and 2 for every stellar model which falls into the rectangle. According to the theory of stellar oscillation, a higher spherical harmonic degree $\ell$ indicates more zones of the stellar surface being divided. Due to the effect of geometrical cancellation, detections of oscillation modes with the higher spherical harmonic degree $\ell$ need much higher precision of the photometric data. Therefore, we follow the work of Chen $\&$ Li (2017), and do not include oscillation frequencies with $\ell \ge$ 3 in this work.

The projected rotational velocity $\upsilon$sin$i$ of IP Per is around 70 km s$^{-1}$ (Miroshnichenko et al. 2001; Zwintz et al. 2014), but the inclination angle $i$ of the stellar rotation axe is unknown. We then consider the equatorial rotation velocity $\upsilon_{\rm e}$ as an adjustable quantity in our calculations. The value of $\upsilon_{\rm e}$ varies from 50 to 180 with a step of 2 in unit of kilometer per second. According to equation (1), every nonradial oscillation mode will split into $2\ell$+1 different frequencies for a given $\upsilon_{\rm e}$. Each oscillation mode with $\ell$ = 1 splits into three different ones, corresponding to $m$ = $-1$, 0, and 1, respectively. Each oscillation mode with $\ell$ = 2 splits into five different ones, corresponding to $m$ = $-2$, -1, 0, 1, and 2, respectively.

To find the best-fitting model to the observations, we perform a $\chi^{2}$ minimization by comparing model frequencies with the observations according to
\begin{equation}
\chi^{2}=\frac{1}{k}\sum(|\nu_{\rm obs, i}-\nu_{\rm mod, i}|^{2}).
\end{equation}
In equation (3), $\nu_{\rm obs, i}$ is the observed frequency, $\nu_{\rm mod, i}$ is the corresponding model frequency, and k is the total number of the observed modes. The observed modes are not identified in advance. Then when we compare frequencies between model and observations, the model frequency nearest to the observed frequency is treated as its possible model counterpart. 

\subsection{Fitting results of IP Per}
Figures 2 and 3 illustrate fitting results of $\chi_{\rm m}^{2}$ as a function of various physical parameters.  In Figures 2 and 3, each circle or square represents one minimum value of $\chi^{2}$ along one evolutionary track, we denote this minimum value with $\chi_{\rm m}^{2}$. Table 2 list 32 candidate models with $\chi_{\rm m}^{2}$ $\le$ 0.10. It can be found in Figures 2 and 3 that these candidate models can be divided into two groups. We distinguish them in Figures 2 and 3 by marking them with rectangles 1 and 2, respectively. The circles falling inside the rectangle 1 correspond to first 20 candidate models in Table 2. The squares falling inside the rectangle 2 correspond to the other 12 candidate models in Table 2. In the following, we will elaborate the details of the fitting results.

Figures 2(a)-(c) present the changes of $\chi_{\rm m}^{2}$ as a function of the equatorial rotation velocity $\upsilon_{\rm e}$, the metallicity $Z$ and the stellar mass $M$, respectively. For candidate models inside the rectangle 1, the free parameters ($\upsilon_{\rm e}$, $Z$, and $M$) converge to $\upsilon_{\rm e} = 91^{+5}_{-3}$ km s$^{-1}$, $Z = 0.009^{+0.004}_{-0.001}$, and $M = 1.64^{+0.10}_{-0.04}$ $M_{\odot}$, respectively. In contrast with the candidate models inside rectangle 1, candidate models in rectangle 2 are rotating faster, metal-richer, and larger. Their values of ($\upsilon_{\rm e}$, $Z$, and $M$) converge to $\upsilon_{\rm e} = 136 \pm 2$ km s$^{-1}$, $Z = 0.021^{+0.001}_{-0.002}$, and $M = 2.00^{+0.02}_{-0.04}$ $M_{\odot}$, respectively. 

Figure 2(d) presents the changes of $\chi_{\rm m}^{2}$ as a function of ages of stars. It can be clearly seen in Figure 2(d) that ages of candidate models inside the rectangle 1 converge to $7.39^{+0.76}_{-0.46}$ Myr, and those inside the rectangle 2 are slightly younger, converging to $6.83 \pm 0.42$ Myr. IP Per is a member star of Per OB2 association. Our asteroseismic ages of IP Per agree well with the results of de Zeeuw et al. (1999) and Bally et al. (2008).

Figure 3(a) presents the changes of $\chi_{\rm m}^{2}$ as a function of the gravitational acceleration $\log g$. In Figure 3(a), we find that the rectangles 1 and 2 overlap with each other.
The surface gravitational accelerations $\log g$ of all candidate models converge to $\log g = 4.041^{+0.009}_{-0.004}$.

Figures 3(b) and 3(c) present the changes of $\chi_{\rm m}^{2}$ as a function of the radius $R$ and the luminosity $\log(L/L_{\odot})$, respectively. The radius $R$ and the luminosity $\log(L/L_{\odot})$ of candidate models inside the rectangle 1 converge to $2.022^{+0.042}_{-0.018}$ $R_{\odot}$ and $1.125^{+0.094}_{-0.046}$, respectively. Candidate models in the rectangle 2 are larger and brighter, converging to $R = 2.235^{+0.008}_{-0.015}$ $R_{\odot}$ and $\log(L/L_{\odot})$ = $1.316\pm0.046$.

Figure 3(d) presents the changes of $\chi_{\rm m}^{2}$ as a function of the acoustic radius $\tau_{0}$, which is defined by Aerts et al. (2010) as
\begin{equation}
\tau_{0}=\int_{0}^{R}\frac{dr}{c_{\rm s}(r)},
\end{equation}
where $c_{\rm s}(r)$ is the adiabatic sound speed. The acoustic radius $\tau_{0}$ is the sound travel time from the surface of the star to the core. As an important asteroseismic quantity, it is usually used to characterize the properties of the stellar envelope (e.g., Ballot et al. 2004; Miglio et al. 2010; Chen et al. 2016, 2017; Chen $\&$ Li 2017, 2018). In Figure 3(d), we find that $\tau_{0}$ converges to $8711^{+68}_{-35}$ s for candidate models inside the rectangle 1, and to $9248^{+39}_{-34}$ s for candidate models inside the rectangle 2.

Figure 3(e) presents the changes of $\chi_{\rm m}^{2}$ as a function of the mass fraction of C12 in the metal composition. Figure 3(f) illustrates the changes of $\chi_{\rm m}^{2}$ as a function of the mass fraction of N14 in the metal composition. In our calculations, we choose the solar metal composition GS98 (Grevesse $\&$ Sauval 1998) as the initial metallicity. The initial mass fractions of C12 and N14 in the metal composition are 0.1721 and 0.0504, respectively. It can be found in Figures 3(e) and 3(f) that the $\delta$ Scuti star IP Per is a pre-main sequence star near the zero-age main sequence. The CN cycle in IP Per has not yet reached the equilibrium state. Based on the candidate models in the rectangel 1, the present mass fractions of C12 and N14 in the convective core are determined to be $0.1478^{+0.0046}_{-0.0020}$ and $0.0787^{+0.0024}_{-0.0054}$ respectively. Fourteen percent of C12 have been turned into N14. However, candidate models in the rectangle 2 suggest that the present mass fractions of C12 and N14 in the convective core are $0.1411^{+0.0013}_{-0.0007}$ and $0.0865^{+0.0008}_{-0.0014}$ respectively. Eighteen percent of C12 have been turned into N14.

Fundamental stellar parameters of the $\delta$ Scuti star IP Per derived from our asteroseismic models are listed in Table 3 . Therein, stellar parameters of Case 1 are obtained on the basis of the candidate models inside the rectangle 1, and those of Case 2 are obtained based on the candidate models inside the rectangle 2. In Case 1, Model 6 has the minimum value of $\chi_{\rm m}^{2} = 0.030$. In Case 2, Model 28 has the minimum value of $\chi_{\rm m}^{2} = 0.068$. We list theoretical frequencies derived from Model 6 and 28 in Table 4, and results of comparisons between model frequencies and the observations in Table 5. According to comparison results of Model 6, $f_{6}$ is identified as a radial mode, $f_{1}$ and $f_{2}$ as two dipole modes, and $f_{3}$, $f_{4}$, $f_{5}$, $f_{7}$, $f_{8}$, and $f_{9}$ as  six quadrupole modes. According to comparison results of Model 28, $f_{4}$ and $f_{7}$ are identified as two radial modes, $f_{1}$, $f_{2}$, $f_{5}$, and $f_{9}$ as four dipole modes, and $f_{3}$, $f_{6}$, and $f_{8}$ as three quadrupole modes.
\subsection{Discussions}
In Section 3.2, our fitting results of IP Per are introduced. The fitting results show that there are two sets of possible parameters for the $\delta$ Scuti star IP Per. The equatorial rotation velocities $\upsilon_{\rm e}$ of the two cases are distinctly different, i.e., $91^{+5}_{-3}$ km s$^{-1}$ for Case 1 and $138 \pm 2$ km s$^{-1}$ for Case 2. The projected rotation velocity $\upsilon\sin i$ is determined to be  $70 \pm 20$ km s$^{-1}$ by  Miroshnichenko et al. (2001) and $70 \pm 4$ km s$^{-1}$ by Zwintz et al. (2014). Based on $\upsilon\sin i =  70 \pm 20$ km s$^{-1}$ of Miroshnichenko et al. (2001),  the inclination angle $i$ is estimated to be $50 \pm 19$ degrees for Case 1, and to be $30 \pm 10$ degrees for Case 2. Based on $\upsilon\sin i =  70 \pm 4$ km s$^{-1}$ of Zwintz et al. (2014), the inclination angle $i$ is suggested to be $50 \pm 7$ degrees for Case 1, and to be $30 \pm 3$ degrees for Case 2. The inclination angle $i$ in Case 2 is much smaller than that in Case 1. Gizon $\&$ Solanki (2003) showed that the ratios of amplitudes for $m$ = $\pm$ 1 and 0 components in a dipole multiplets and those of $m$ = $\pm$ 2, $\pm$ 1 and 0 components in a quadrupole multiplets exhibit different dependences on $i$. Unfortunately, it can be noticed in Table 5 that none of complete mutiplets is found in oscillation spectrum of IP Per.

Besides, we notice in Table 3 that the effective temperature $T_{\rm eff}$ of the two cases are slightly different, i.e., $7766^{+348}_{-173}$ K for Case 1 and $8243^{+211}_{-185}$ K for Case 2. Based on spectroscopic analyses, Miroshnichenko et al. (2001) obtained that $T_{\rm eff}$ of IP Per is $7960 \pm 50$ K, and Zwintz et al. (2014) suggested that IP Per is a star with $T_{\rm eff} = 8000 \pm 100$ K. 
The asteroseismic parameters of Case 1 are in agreement with both the results of Miroshnichenko et al. (2001) and Zwintz et al. (2014), and those of Case 2 agree with the results of Zwintz et al. (2014).

In Table 3, we find that candidate models of Case 1 are metal-deficient, and their metallicities converge to $Z = 0.009^{+0.004}_{-0.001}$. While metallicities of candidate models in Case 2 converge to $Z = 0.021^{+0.001}_{-0.002}$, similar to the solar metallicity. Miroshnichenko et al. (2001) found that [M/H] of IP Per is $-0.41 \pm 0.1$. Zwintz et al. (2014) estimated [Fe/H] of IP Per to be $-0.24$. Both results of Miroshnichenko et al. (2001) and Zwintz et al. (2014) agree well with our asteroseismic result in Case 1. 

Moreover, IP Per is a member star of Per OB2 association, which is at a distance of D = $296\pm17$ pc (de Zeeuw et al. 1999). Besides, the parallax of IP Per from Gaia R2 is $3.2498 \pm 0.0777$ mas (Gaia Collaboration 2018), corresponding to the distance of D = $308 \pm 8$ pc. The two distances agree well with each other. Then the luminosity of IP Per can be estimated according to the following expression
\begin{equation}
\log(L/L_{\odot}) = 0.4[4.75 - V + A_{V} -  BC(V) + (m -M )_{0}]
\end{equation}
(Wu et al. 2014), where the distance modulus $(m- M)_{0}$ can be deduced according to 
\begin{equation}
 (m -M )_{0} = 5\log D -5.
\end{equation}
The apparent magnitude of IP Per in Johnson V is 10.374 mag (Ripepi et al. 2006). $\check{\rm C}$ernis (1993) investigated the extinction in the direction of Per OB2 association and obtained that the mean extinction of the Per OB2 association stars is $0.95 \pm 0.08$ mag. Manoj et al. (2006) estimated the extinction of IP Per to be 0.76 mag, slightly smaller than that of $\check{\rm C}$ernis (1993). IP Per is a pre-main sequence star, we thus use the bolometric corrections BC(V) = $-0.12$ from Kenyon $\&$ Hartmann (1995). According to Equation (5), the luminosity of IP Per is estimated to be $\log(L/L_{\odot}) = 1.16 \pm 0.06$ with the extinction of $\check{\rm C}$ernis (1993), and to be $\log(L/L_{\odot}) = 1.08 \pm 0.02$ with the extinction of Manoj et al. (2006). Both of them are in agreement with our asteroseismic result in Case 1. 

In addition, it can be clearly seen in Figure 3(a) that $\log g$ of all of our candidate models converge to $4.037-4.049$. However, Miroshnichenko et al. (2001) and Zwintz et al. (2014) suggested that the surface gravitational acceleration $\log g$ of IP Per is around 4.4. Our asteroseismic result is significantly lower than the results of Miroshnichenko et al. (2001) and Zwintz et al. (2014).

Based on the above considerations, our asteroseismic parameters of Case 1 agree well with availably photometric and spectroscopic measurements. Thus we adopt Model 6 as the best-fitting model, which is marked with a crossing in Figure 4. Fundamental stellar parameters of IP Per are suggested to be $\upsilon_{\rm e} = 91 ^{+5}_{-3}$ km s$^{-1}$, $Z = 0.009^{+0.004}_{-0.001}$, $M = 1.64^{+0.10}_{-0.04}$ $M_{\odot}$, $T_{\rm eff} = 7766^{+348}_{-173}$ K, $\log(L/L_{\odot}) = 1.125 ^{+0.094}_{-0.046}$, $\log g = 4.041 ^{+0.008}_{-0.003}$, $R = 2.022 ^{+0.042}_{-0.018}$ $R_{\odot}$,  $\tau_{0} = 8711^{+68}_{-35}$ s, age = 7.39$^{+0.76}_{-0.46}$ Myr.

\section{ Summary and Conclusions}
In this work, we have performed numerical calculations and asteroseismic analyses for the $\delta$ Scuti  pulsator IP Per in detail. The main results of our work are summarized  as follows:

1.Our fitting results show that there are two sets of theoretical models that could reproduced the observed frequencies. In consideration of other available photometric and spectroscopic measurements, our best fitting stellar parameters for IP Per are $Z=0.009^{+0.004}_{-0.001}$, $M = 1.64^{+0.10}_{-0.04}$ $M_{\odot}$, $T_{\rm eff} = 7766^{+348}_{-173}$ K, $\log L/L_{\odot} = 1.125^{+0.094}_{-0.046}$, $\log g = 4.041^{+0.008}_{-0.003}$ dex, $R = 2.022^{+0.042}_{-0.018}$ $R_{\odot}$,  $\tau_{0} = 8711^{+68}_{-35}$ s, age = $7.39^{+0.76}_{-0.46}$ Myr (i.e., our Case 1). 

2.The equatorial rotation velocity $\upsilon_{\rm e}$ is found to converge to $91^{+5}_{-3}$ km s$^{-1}$. The projected rotational velocity $\upsilon\sin i$  of IP Per is determined to be $70\pm20$ km s$^{-1}$ by Miroshnichenko et al. (2001) and $70\pm4$ km s$^{-1}$ by Zwintz et al. (2014). Hence  the inclination angle $i$ is suggested to be $50\pm19$ degrees and $50\pm7$ degrees respectively.

3.According to our calculations, IP Per is found to be a pre-main sequence star where CN cylce has not arrived at the equilibrium state. The present mass fraction of C12 in metal is $0.1478^{+0.0046}_{-0.0020}$, and that of N14 in metal is $0.0787^{+0.0024}_{-0.0054}$. Fourteen percent of C12 have turned into N14.

4.Based on our best-fitting model, $f_{6}$ is suggested to be a radial mode, $f_{1}$ and $f_{2}$ to be two dipole modes, and $f_{3}$, $f_{4}$, $f_{5}$, $f_{7}$, $f_{8}$, and $f_{9}$ to be six quadrupole modes. 

\acknowledgments
This work is funded by the NSFC of China (Grant No. 11333006, 11521303, 11803082) and by the foundation of Light of West China Program from Chinese Academy of Sciences. We are sincerely grateful to an anonymous referee for instructive advice and productive suggestions. The authors gratefully acknowledge the computing time granted by the Yunnan Observatories, and provided on the facilities at the Yunnan Observatories Supercomputing Platform.

\begin{figure*}
 \epsscale{1.1}
 \centering
 \plotone{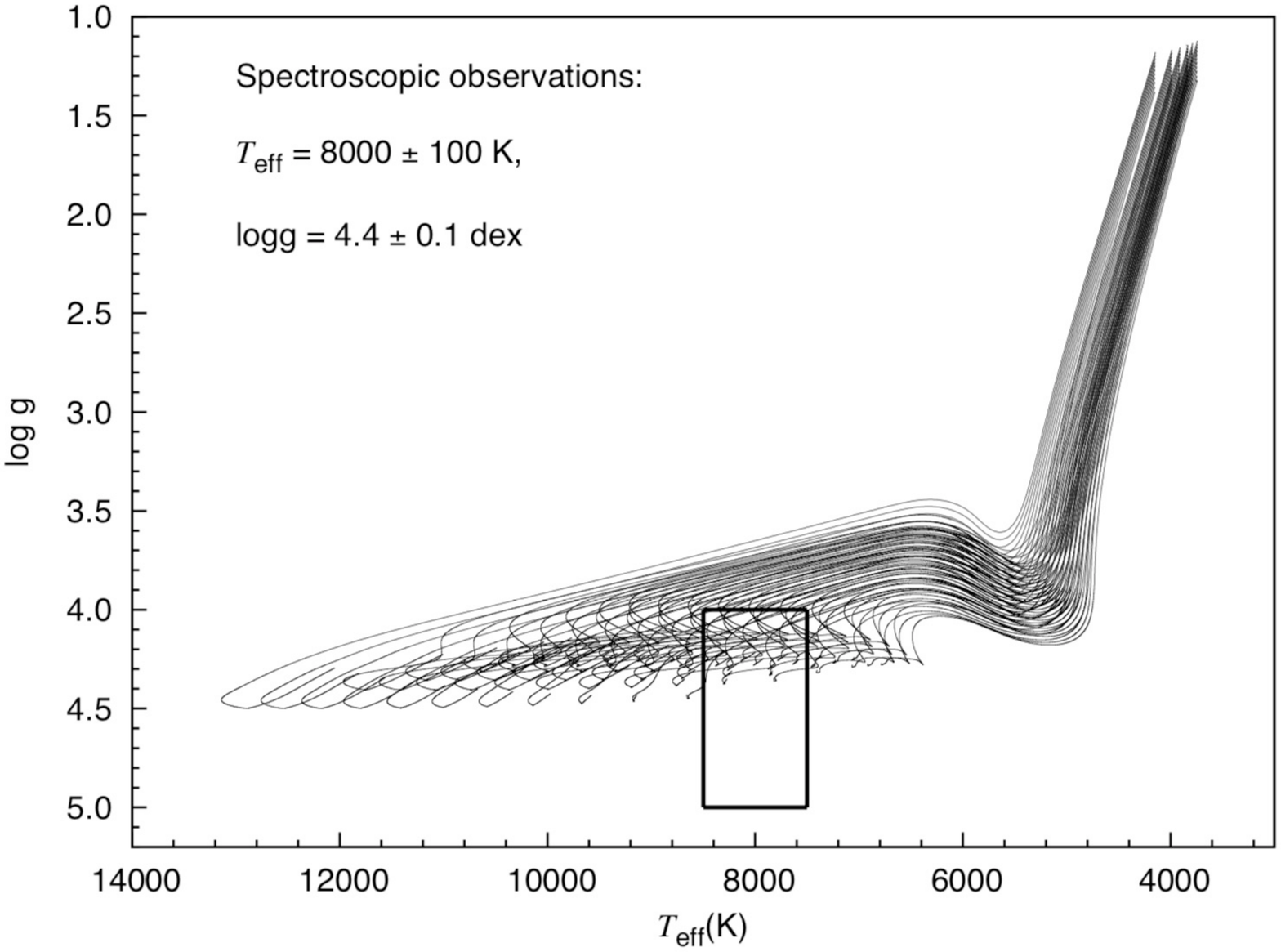}
  \caption{Visualisation of the model calculations for IP Per. In this figure, the mass $M$ varies from 1.50 $M_{\odot}$ to 2.50 $M_{\odot}$ with a step of 0.1 $M_{\odot}$, and the metallicity Z varies from 0.005 to 0.030 with a step of 0.005. The rectangle corresponds to the range:4.0 $<$ $\log g$ $<$ 5.0 and 7500 K $<$ $T_{\rm eff}$ $<$ 8500 K.}
  \label{Figure.1}
\end{figure*}

\begin{figure*}
  \epsscale{1.2}
  \centering
 \plotone{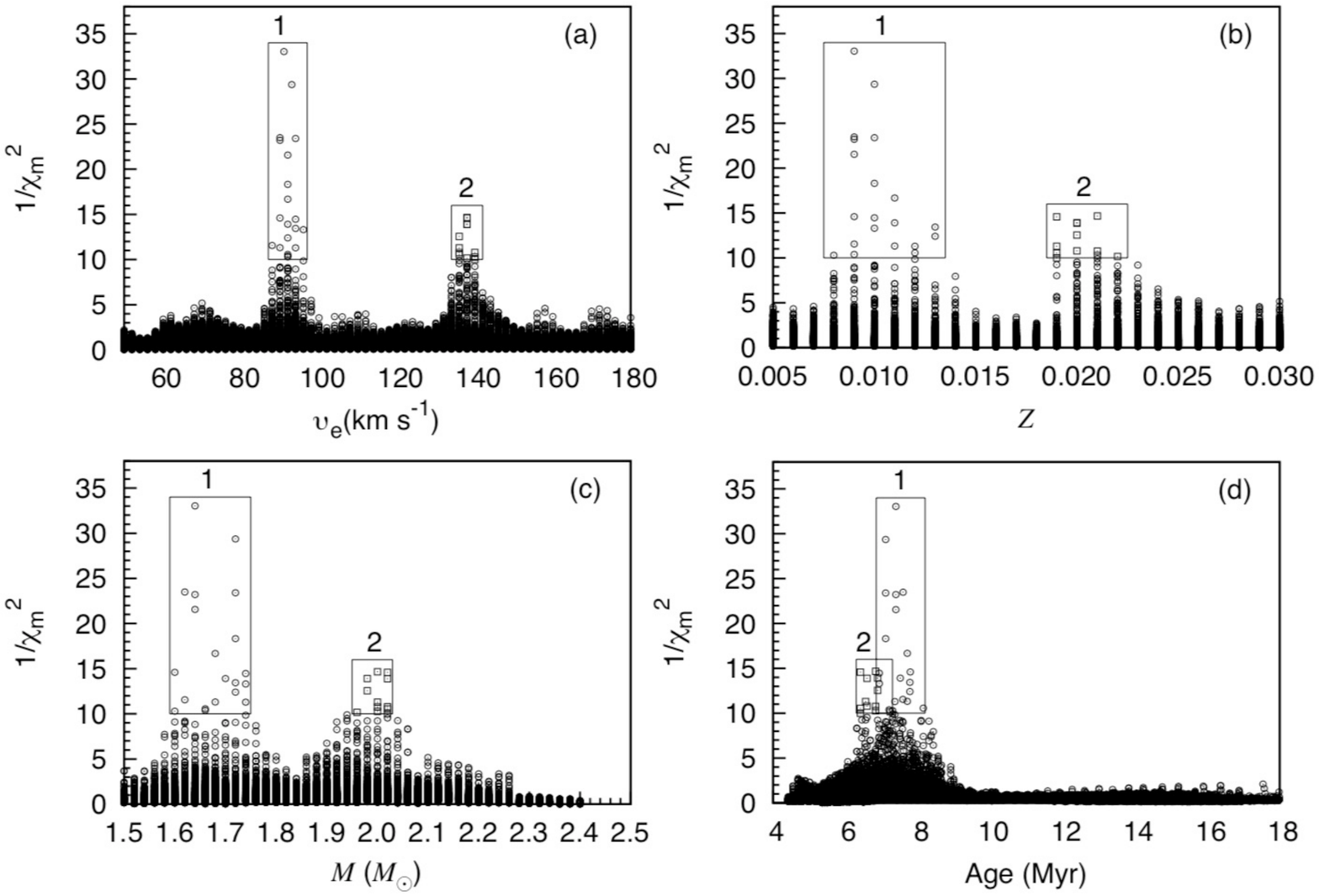}
  \caption{Visualisation of the resulting 1/$\chi_{\rm m}^{2}$ as a function of he equatorial rotation velocity $\upsilon_{\rm e}$ (Panel (a)), the metallicity $Z$ (Panel (b)), the stellar mass $M$ (Panel (c)), and the age (Panel (d)), respectively. The circles in rectangle 1 correspond to the first 20 candidate models of Table 2, and the squares in rectangle 2 correspond to the other 12 candidate models of Table 2. }
  \label{Figure.2}
\end{figure*}

\begin{figure*}
  \epsscale{1.1}
   \centering
  \plotone{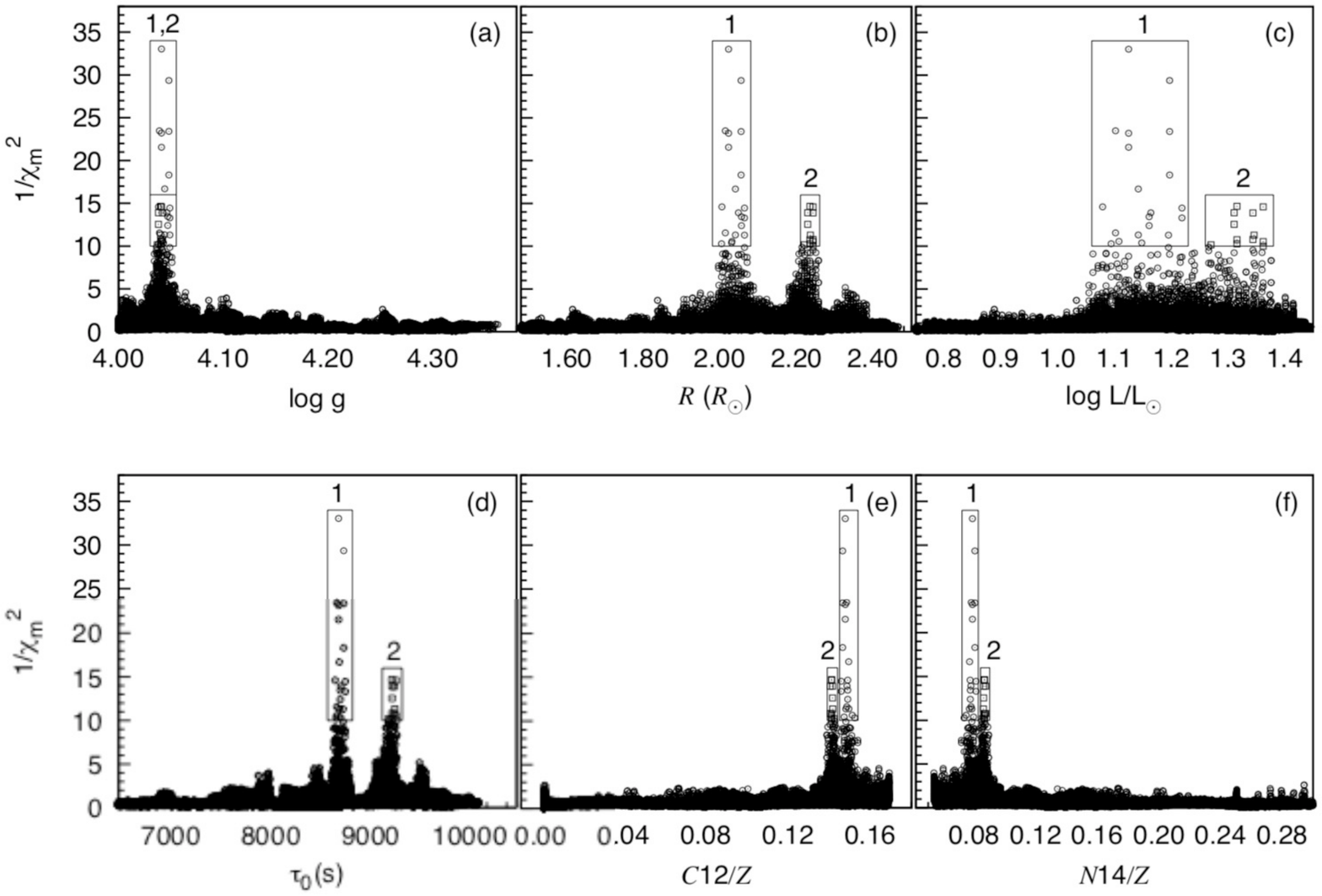}
  \caption{Visualisation of the resulting 1/$\chi_{\rm m}^{2}$ as a function of the gravitational acceleration $\log g$ (Panel (a)), the radius R (Panel (b)), the luminosity log$L/L_{\odot}$ (Panel (c)), the acoustic radius $\tau_{0}$ (Panel (d)), the mass fraction of C12 in metal composition C12/$Z$ (Panel (e)), and the mass fraction of N14 in metal composition N14/$Z$ (Panel (f)), respectively. The circles in rectangle 1 correspond to the first 20 candidate models of Table 2, and the squares in rectangle 2 correspond to the other 12 candidate models of Table 2. }
  \label{Figure.3}
\end{figure*}

\begin{figure*}
 \epsscale{1.0}
 \centering
 \plotone{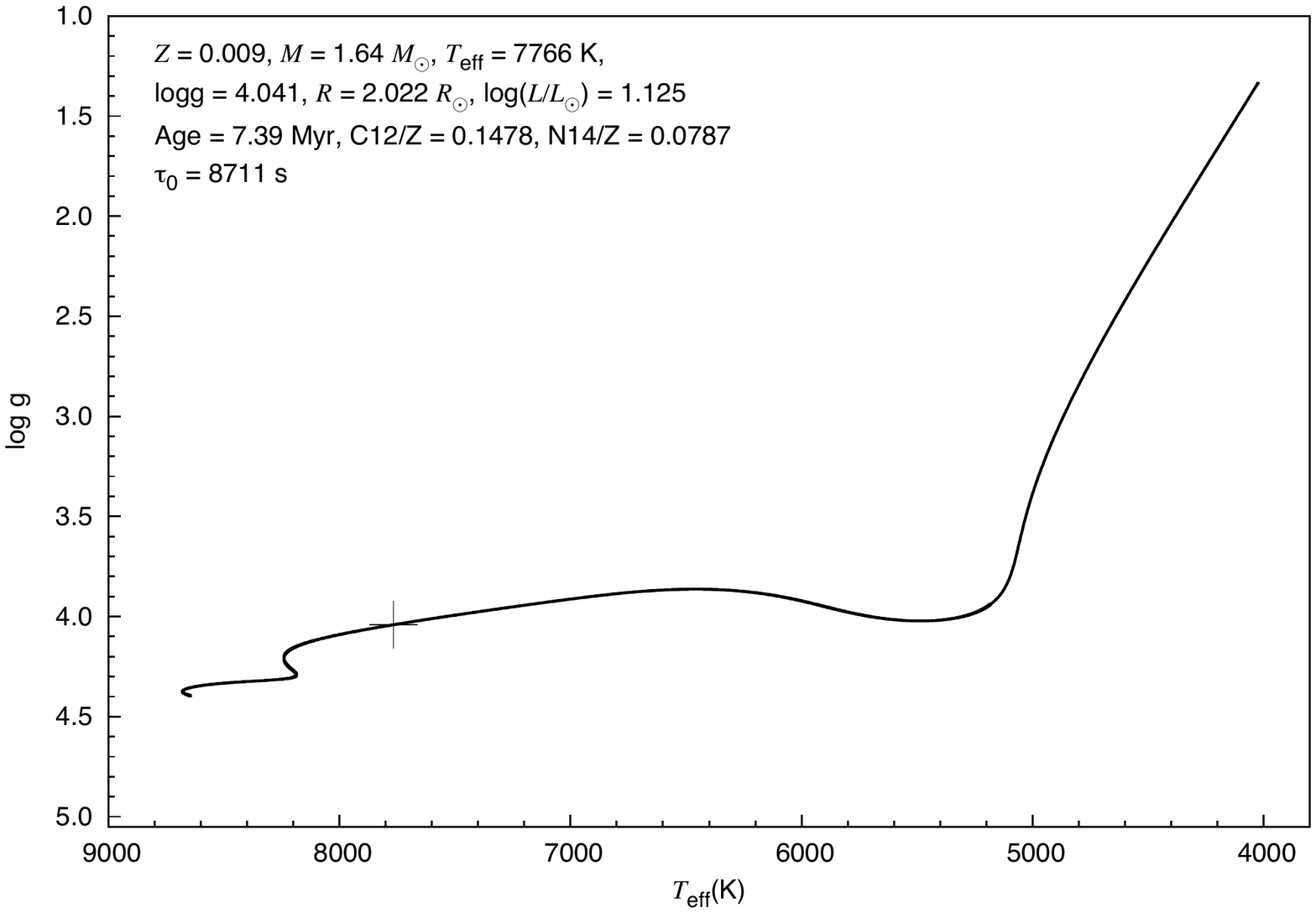}
  \caption{Visualisation of the evolutionary track of the star with $Z$ = 0.009 and $M$ = 1.64 $M_{\odot}$. The crossing denotes the best-fitting model (Model 6), whose physical parameters are labelled in the top of the figure.}
  \label{Figure.1}
\end{figure*}

\begin{table*}
\LARGE
\centering
\caption{\label{t1} Nine independent frequencies of IP Per obtained by Ripepi et al. (2006). The first column is the serial number of the observed frequencies. The second and third columns are observed frequencies in unit of cycle per day and $\mu$Hz, respectively. The values in the fourth column are amplitudes of the observed frequency in unit of mmag. The uncertainty of the observed frequency is 0.11 in unit of cycle per day and 1.273 in unit of $\mu$Hz.}
\begin{tabular}{lccccc}
\hline\hline
ID            &Freq.          &Freq.               &Ampl.    \\
               &(d$^{-1}$)     &($\mu$Hz)     &(mmag)    \\
\hline
$f_{1}$   &9.30         &107.64         &1.3   \\
$f_{2}$   &22.89      &264.93         &1.9    \\
$f_{3}$   &23.99      &277.66         &1.3    \\
$f_{4}$   &28.79      &333.22         &1.5    \\
$f_{5}$   &30.45      &352.43        &1.8    \\
$f_{6}$   &34.60      &400.46        &1.5    \\
$f_{7}$   &41.11        &475.81         &1.2    \\
$f_{8}$   &48.23      &558.22         &1.6   \\
$f_{9}$   &52.00      &601.85          &1.1    \\
\hline
\end{tabular}
\end{table*}

\begin{table*}
\centering
\caption{\label{t2}Candidate models with $\chi_{\rm m}^{2}$ $<$ 0.10. 
The column named Model is the serial number of candidate models. $\upsilon_{\rm eq}$ is the rotational velocity at the equator of the star.  $\tau_{0}$ is the acoustic radius defined as equation (4). $X_{\rm c}$ is the mass fraction of hydrogen in the center of the star. C12/Z is the mass fraction of C12 in the metal. N14/Z is the mass fraction of N14 in the metal.}
\begin{tabular}{ccccccccccccccccc}
\hline\hline
Model &$\upsilon_{\rm eq}$ &$Z$  &$M$   &$T_{\rm eff}$ &log($L/L_{\odot}$)   &$R$ &logg  &$\tau_{0}$  &$X_{\rm c}$ &$C12/Z$  &$N14/Z$ &Age &$\chi_{\rm m}^{2}$  \\
&(km s$^{-1}$)  &     &($M_{\odot}$) &(K) &  &$(R_{\odot})$  &(dex)  &$(\rm s)$ & & & &(Myr)  &  \\
\hline
1  & 88 &0.009 &1.62 &7679 &1.102 &2.014 &4.039 &8694 &0.7320 &0.1485 &0.0779 & 7.59 &0.087\\
2  & 90 &0.009 &1.60 &7593 &1.079 &2.005 &4.038 &8676 &0.7320 &0.1490 &0.0773 & 7.79 &0.069\\
3  & 90 &0.009 &1.62 &7680 &1.102 &2.014 &4.039 &8693 &0.7320 &0.1484 &0.0780 & 7.59 &0.043\\
4  & 90 &0.009 &1.64 &7766 &1.125 &2.022 &4.041 &8711 &0.7320 &0.1478 &0.0787 & 7.39 &0.043\\
5  & 90 &0.011 &1.68 &7810 &1.142 &2.040 &4.044 &8720 &0.7270 &0.1495 &0.0768 & 7.71 &0.088\\
6  & 91 &0.009 &1.64 &7766 &1.125 &2.022 &4.041 &8711 &0.7320 &0.1478 &0.0787 & 7.39 &0.030\\
7  & 92 &0.008 &1.60 &7665 &1.094 &2.004 &4.038 &8688 &0.7346 &0.1524 &0.0733 & 7.30 &0.097\\
8  & 92 &0.009 &1.64 &7766 &1.125 &2.022 &4.041 &8711 &0.7320 &0.1478 &0.0787 & 7.39 &0.046\\
9  & 92 &0.009 &1.66 &7853 &1.148 &2.030 &4.043 &8727 &0.7320 &0.1472 &0.0795 & 7.20 &0.096\\
10 & 92 &0.010 &1.72 &8031 &1.197 &2.055 &4.048 &8763 &0.7295 &0.1465 &0.0803 & 7.11 &0.055\\
11 & 92 &0.011 &1.68 &7810 &1.142 &2.040 &4.044 &8718 &0.7270 &0.1494 &0.0769 & 7.71 &0.060\\
12 & 92 &0.011 &1.70 &7894 &1.164 &2.048 &4.046 &8734 &0.7270 &0.1486 &0.0778 & 7.52 &0.072\\
13 & 92 &0.012 &1.66 &7674 &1.108 &2.031 &4.042 &8691 &0.7244 &0.1512 &0.0747 & 8.15 &0.095\\
14 & 92 &0.013 &1.72 &7864 &1.161 &2.056 &4.047 &8728 &0.7219 &0.1490 &0.0774 & 7.78 &0.081\\
15 & 93 &0.010 &1.72 &8031 &1.197 &2.055 &4.048 &8763 &0.7295 &0.1465 &0.0803 & 7.11 &0.034\\
16 & 94 &0.010 &1.72 &8031 &1.197 &2.055 &4.048 &8763 &0.7295 &0.1465 &0.0803 & 7.11 &0.043\\
17 & 94 &0.010 &1.74 &8114 &1.219 &2.064 &4.049 &8779 &0.7295 &0.1458 &0.0811 & 6.93 &0.069\\
18 & 94 &0.012 &1.74 &8004 &1.195 &2.064 &4.049 &8757 &0.7244 &0.1481 &0.0784 & 7.36 &0.089\\
19 & 94 &0.013 &1.72 &7864 &1.161 &2.056 &4.047 &8728 &0.7219 &0.1490 &0.0774 & 7.78 &0.074\\
20 & 96 &0.010 &1.74 &8114 &1.219 &2.064 &4.049 &8779 &0.7295 &0.1458 &0.0811 & 6.93 &0.075\\
21 &136 &0.019 &2.00 &8387 &1.346 &2.236 &4.040 &9276 &0.7066 &0.1417 &0.0859 & 6.56 &0.089\\
22 &136 &0.019 &2.02 &8454 &1.362 &2.243 &4.041 &9287 &0.7066 &0.1411 &0.0866 & 6.41 &0.095\\
23 &136 &0.020 &1.98 &8235 &1.311 &2.229 &4.038 &9248 &0.7041 &0.1416 &0.0860 & 6.89 &0.080\\
24 &136 &0.021 &2.00 &8243 &1.316 &2.235 &4.040 &9248 &0.7015 &0.1411 &0.0865 & 6.83 &0.093\\
25 &138 &0.019 &2.02 &8454 &1.362 &2.243 &4.041 &9287 &0.7066 &0.1411 &0.0866 & 6.41 &0.069\\
26 &138 &0.020 &1.98 &8235 &1.311 &2.229 &4.038 &9248 &0.7041 &0.1416 &0.0860 & 6.89 &0.072\\
27 &138 &0.020 &2.02 &8366 &1.344 &2.243 &4.042 &9269 &0.7041 &0.1404 &0.0873 & 6.59 &0.072\\
28 &138 &0.021 &2.00 &8243 &1.316 &2.235 &4.040 &9248 &0.7015 &0.1411 &0.0865 & 6.83 &0.068\\
29 &138 &0.022 &1.96 &8058 &1.270 &2.220 &4.037 &9214 &0.6990 &0.1424 &0.0851 & 7.24 &0.098\\
30 &140 &0.019 &2.02 &8454 &1.362 &2.243 &4.041 &9287 &0.7066 &0.1411 &0.0866 & 6.41 &0.100\\
31 &140 &0.020 &2.02 &8366 &1.344 &2.243 &4.042 &9269 &0.7041 &0.1404 &0.0873 & 6.59 &0.093\\
32 &140 &0.021 &2.00 &8243 &1.316 &2.235 &4.040 &9248 &0.7015 &0.1411 &0.0865 & 6.83 &0.098\\
\hline
\end{tabular}
\end{table*}

\begin{table*}
\large
\centering
\caption{\label{t3}Fundamental parameters of the $\delta$ Scuti star IP Per derived from our asteroseismic models. Parameters in the second column named Case 1 derived from candidate models in rectangle 1. Parameters in the third column named Case 2 derived from candidate models in rectangle 2. The fourth column lists the observed parameters from photometry and spectroscopy.}
\begin{tabular}{llll}
\hline\hline
Parameter                                               &Case 1                              &Case 2             &Other works\\
\hline
$\upsilon_{\rm eq}$(km s$^{-1}$)        &$91^{+5}_{-3}$                &$138\pm2$   & 
$\upsilon\sin i = 70 \pm 20$(Miroshnichenko et al. 2001)\\
&&&$\upsilon\sin i = 70 \pm 4$(Zwintz et al. 2014)\\
$M(M_{\odot})$                                     &$1.64^{+0.10}_{-0.04}$  &$2.00^{+0.02}_{-0.04}$ \\
$Z$                                                        &$0.009^{+0.004}_{-0.001}$      &$0.021^{+0.001}_{-0.002}$&[M/H] = $-0.41\pm 0.1$ (Miroshnichenko et al. 2001) \\
&&&[Fe/H] = $-0.24$ (Zwintz et al. 2014)\\
$T_{\rm eff}$(K)                                   &$7766^{+348}_{-173}$               &$8243^{+211}_{-185}$ 
&$7950\pm50$ (Miroshnichenko et al. 2001)\\
&&&$8000\pm100$ (Zwintz et al. 2014)\\
log$g$ (dex)                                         &$4.041^{+0.008}_{-0.003}$         &$4.040^{+0.002}_{-0.003}$ &$4.38\pm0.04$ (Miroshnichenko et al. 2001)\\
&&&$4.4\pm0.1$ (Zwintz et al. 2014)\\
$R(R_{\odot})$                                      &$2.022^{+0.042}_{-0.018}$         &$2.235^{+0.008}_{-0.015}$ \\
log($L/L_{\odot}$)                                &$1.125^{+0.094}_{-0.046}$           &$1.316\pm0.046$ 
&$1.16\pm0.06$ or $1.08\pm0.02$ (This work) \\
Age (Myr)                                              &$7.39^{+0.76}_{-0.46}$                  &$6.83\pm0.42$
& $<$ 15 Myr (de Zeeuw et al. 1999; Bally et al. 2008)\\
$\tau_{0}$(s)                                        &$8711^{+68}_{-35}$                        &$9248^{+39}_{-34}$ \\
$C12/Z$                                                &$0.1478^{+0.0046}_{-0.0020}$    &$0.1411^{+0.0013}_{-0.0007}$ \\
$N14/Z$                                               &$0.0787^{+0.0024}_{-0.0054}$    &$0.0865^{+0.0008}_{-0.0014}$ \\
\hline
\end{tabular}
\end{table*}

\begin{table*}
\centering
\caption{\label{t4}Theoretical calculated frequencies derived from Model 6 and Model 28. $\nu_{\rm mod}$ is the model frequency, $\ell$ and $n$ are its corresponding spherical harmonic degree and radial order. $\beta_{\ell,n}$ is the rotational parameter determining the size of the frequency splitting, and $\beta_{\ell,n}$ can be described quantitatively by equation (2).}
\label{observed frequencies}
\begin{tabular}{ccccccccccc}
\hline\hline
Model &$\nu_{\rm mod}$ &$(\ell,n)$ &$\beta_{\ell, n}$ &$\nu_{\rm mod}$&$(\ell,n)$ &$\beta_{\ell, n}$ &$\nu_{\rm mod}$&$(\ell,n)$ &$\beta_{\ell, n}$\\
&($\mu$Hz)   & & &($\mu$Hz)   & &&($\mu$Hz)\\
\hline
& 160.033     &(0,  0)&         & 145.012  &(1, 0)  &0.529 &102.809 &(2,  -3)&0.835&\\
& 206.662    &(0,   1)&         & 164.845 &(1, 1)  &0.958 &123.539  &(2,  -2)&0.816&\\
& 255.132     &(0,  2)&         & 213.151  &(1,  2) &0.996 & 149.476 &(2,  -1)&0.836&\\
& 303.188    &(0,   3)&         &264.826 &(1,  3) &0.999 & 169.705 &(2,   0)&0.962&\\
& 351.615     &(0,   4)&         & 317.296 &(1, 4)  &0.998 & 197.723 &(2,   0)&0.856&\\
& 400.601    &(0,   5)&         & 369.911 &(1, 5)  &0.996 & 227.546 &(2,   1)&0.947&\\
& 450.015    &(0,   6)&         & 421.471  &(1, 6) &0.993 & 258.387 &(2,   2)&0.942&\\
6& 500.438 &(0,   7)&          &472.586 &(1, 7)  &0.992 & 295.151 &(2,    3)&0.877&\\
& 552.561    &(0,   8)&          &524.632 &(1, 8) &0.991 & 342.795 &(2,    4)&0.917&\\
& 606.245   &(0,   9)&          &578.026 &(1, 9)  &0.990 & 393.546 &(2,   5)&0.945&\\
& 661.093   &(0,  10)&          &632.584 &(1,10) &0.989 & 444.350 &(2,   6)&0.960&\\
&                  &           &           &687.962 &(1,11)  &0.989 & 495.546 &(2,  7)&0.969&\\
& 63.687     &(1,  -3)&0.510 &               &           &           & 548.099 &(2,  8)&0.974&\\
& 78.592     &(1,  -2)&0.503 &  75.967 &(2,-5) &0.847 & 602.006 &(2,  9)&0.978&\\
& 102.220   &(1,   -1)&0.497 &  87.507 &(2,-4) &0.844 & 656.948 &(2,  10)&0.981&\\
\hline
& 150.719      &(0,  0)&          & 114.373&(1,  0)&0.483& 120.453&(2,      -1)&0.795&\\
& 195.079     &(0,   1)&         & 154.793&(1,   1)&0.983& 150.013&(2,       0)&0.930&\\
& 240.902    &(0,   2)&         & 201.284&(1,  2)&0.996& 175.035&(2,       0)&0.849&\\
& 286.512    &(0,   3)&         & 250.677&(1,  3)&0.998& 206.810&(2,       1)&0.996&\\
& 332.648    &(0,  4)&         & 301.189&(1,   4)&0.996& 238.786&(2,       2)&0.891&\\
& 379.991     &(0,  5)&        & 352.061&(1,   5)&0.993& 277.610&(2,       3)&0.875&\\
28& 427.642&(0,  6)&        & 401.874&(1,   6)&0.991& 324.430&(2,       4)&0.921&\\
& 475.799    &(0,  7)&        & 450.869&(1,   7)&0.989& 373.881&(2,       5)&0.949&\\
& 525.254   &(0,  8)&         & 500.261&(1,   8)&0.988& 422.906&(2,       6)&0.964&\\
& 575.921(   &0,  9)&         & 550.670&(1,   9)&0.987& 471.756&(2,       7)&0.971&\\
& 627.566   &(0, 10)&        & 601.945&(1,  10)&0.987& 521.520&(2,       8)&0.976&\\
& 680.018   &(0,  11)&       & 653.983&(1,  11)&0.987& 572.296&(2,       9)&0.980&\\
&                  &           &       &                 &         &          & 623.917&(2,      10)&0.982&\\
&  76.286   &(1,  -1)&0.495&  93.536&(2,  -2)&0.834& 676.262&(2,      11)&0.984&\\
\hline
\end{tabular}
\end{table*}

\begin{table*}
\centering
\large
\caption{\label{t5}Comparisons between model frequencies of Model 6 and Model 28 and observations. $\nu_{\rm obs}$ is the observed frequency. $\sigma_{\rm obs}$ is the uncertainty of the observed frequency. $\nu_{\rm mod}$ is the model frequency. ($\ell, n, m$) are the spherical harmonic degree, the radial order, and the azimuthal number of the model frequency, respectively. $|$$\nu_{\rm obs}$$-$$\nu_{\rm mod}$$|$ is the difference  between the observed frequency and its model counterpart.}
\begin{tabular}{ccccclc}
\hline\hline
Model  &ID  &$\nu_{\rm obs}$ &$\sigma_{\rm obs}$  &$\nu_{\rm mod}$ &($\ell$, $n$, $m$) & $|\nu_{\rm obs}-\nu_{\rm mod}|$\\
            &       &($\mu$Hz)       &($\mu$Hz)    &($\mu$Hz)              &                               &($\mu$Hz)\\
\hline
&$f_{1}$    &107.64    &1.273        &107.335          &(1,-1,+1)          &0.305\\
&$f_{2}$    &264.93   &1.273        &264.826         &(1, 3, 0)          &0.104\\
&$f_{3}$    &277.66   &1.273        &277.775          &(2, 2, 2)          &0.115\\
&$f_{4}$    &333.22   &1.273        &333.358         &(2, 4,-1)        &0.138\\
6&$f_{5}$  &352.43   &1.273        &352.231         &(2, 4,+1)        &0.199\\
&$f_{6}$    &400.46   &1.273        &400.601         &(0, 5, 0)         &0.141\\
&$f_{7}$    &475.81    &1.273        &475.602           &(2, 7,-2)       &0.208\\
&$f_{8}$    &558.22   &1.273        &558.122          &(2, 8,+1)         &0.098\\
&$f_{9}$    &601.85   &1.273        &602.006          &(2, 9, 0)          &0.156\\
\hline
&$f_{1}$    &107.64    &1.273       &107.554         &(1, 0,-1)          &0.086\\
&$f_{2}$    &264.93   &1.273       &264.776         &(1, 3,+1)         &0.164\\
&$f_{3}$    &277.66   &1.273        &277.610         &(2, 3, 0)          &0.050\\
&$f_{4}$    &333.22   &1.273        &332.648        &(0, 4, 0)         &0.572\\
28&$f_{5}$&352.43   &1.273       &352.061         &(1, 5, 0)          &0.369\\
&$f_{6}$    &400.46   &1.273       &400.675         &(2, 5,+2)       &0.215\\
&$f_{7}$    &475.81    &1.273        &475.799        &(0, 7, 0)          &0.011\\
&$f_{8}$    &558.22   &1.273        &558.461        &(2, 9,-1)         &0.241\\
&$f_{9}$    &601.85   &1.273         &601.945        &(1, 10,0)          &0.095\\
\hline
\end{tabular}
\end{table*}

\end{document}